\documentclass[cameraready]{Interspeech}

\title{Towards Data-free and Training-free Compression for Speech Foundation Models Using Parameter Clustering}

\author[affiliation={1}]{Haoning}{Xu}
\author[affiliation={1}]{Zhaoqing}{Li}
\author[affiliation={1}]{Huimeng}{Wang}
\author[affiliation={1}]{Youjun}{Chen}
\author[affiliation={1}]{Chengxi}{Deng}
\author[affiliation={2}]{Mengzhe}{Geng}
\author[affiliation={1}]{Xunying}{Liu}

\address{$^1$ The Chinese University of Hong Kong, Hong Kong SAR, China \\
$^2$ National Research Council Canada, Canada }
\email{hnxu@se.cuhk.edu.hk, xyliu@se.cuhk.edu.hk}

\keywords{speech recognition, parameter clustering, model pruning, speech foundation models}

\usepackage{comment}

\usepackage{caption}      
\usepackage{hyperref}     

\usepackage{amsmath}      
\usepackage{amssymb}    

\usepackage{amsfonts}     
\usepackage{algorithmic}  
\usepackage{algorithm}    
\usepackage{array}        
\usepackage{multirow}     
\usepackage{booktabs}     
\usepackage{graphicx}     
\usepackage[caption=false,font=normalsize,labelfont=sf,textfont=sf]{subfig}  
\usepackage{stfloats}     
\usepackage{float}        
\usepackage{textcomp}     
\usepackage{verbatim}     
\usepackage{url}          
\usepackage{bbding}       
\usepackage{pifont}       
\usepackage{enumitem}     
\usepackage{stackengine}  
\usepackage[table]{xcolor}

\newcommand{\tim}[1]{{\color{black}{#1}}}
\newcommand{\cmark}{\ding{51}}
\newcommand{\xmark}{\ding{55}}
\newcommand{\zq}[1]{{\color{black}{#1}}}
\newcommand{\abswer}[1]{\rlap{\hspace{0.1em}\scriptsize \scalebox{0.9}{\underline{#1}}}}
\newcommand{\abswerblue}[1]{\rlap{\hspace{0.1em}\scriptsize \scalebox{0.9}{\textcolor{red}{\underline{#1}}}}}

\begin{document}

\maketitle

\begin{abstract}

\zq{This paper presents a novel data-free and training-free compression approach for speech foundation models using channel-wise clustering via k-means.
More fine-grained, mixed sparsity pruning by layer-level varying number of parameter clusters is also explored. 
Experiments conducted on the LibriSpeech dataset suggest that when operating with pruning sparsity of 50\% on HuBERT-large, consistent WER reductions of 27.73\%/18.61\% absolute (34.37\%/21.91\% relative) over the magnitude-based pruning were obtained on the test-clean and test-other subsets before fine-tuning and 0.19\%/0.79\% absolute (3.36\%/4.62\% relative) after fine-tuning with only 3 epochs.
Similar WER reductions of 2.86\%/5.02\% absolute (59.21\%/55.29\% relative) were observed against magnitude-based pruning on Whisper-large-v3 at 10\% sparsity, all with no significant WER increase relative to the uncompressed baseline.
}

\end{abstract}
\vspace{-3mm}
\section{Introduction}
\vspace{-1mm}

\label{sec:intro}
Recent advances in speech technology have been driven by speech foundation models, including self-supervised learning (SSL) models such as wav2vec2.0~\cite{baevski2020wav2vec}, HuBERT~\cite{hsu2021HuBERT} and WavLM~\cite{chen2022wavlm}, as well as the supervised learning models such as Whisper~\cite{openai2022whisper}, all of which significantly boost automatic speech recognition (ASR) performance. Despite these advances, the widespread use of these models in resource-constrained scenarios (e.g., on-device environments) remains limited due to their large memory footprint and high computational demands.

To tackle this challenge, numerous studies have investigated various neural network compression techniques for ASR tasks, such as (but not restricted to): \textbf{1) low-bit quantization} techniques, which decrease memory usage by substituting floating-point weights with low-precision values~\cite{uq-2bcfm,uq-4bcfm,ibert,xu2025effective,mq-person,li25_interspeech}; and \textbf{2) architectural compression} methods, which focus on mitigating structural redundancy in models, including low-rank matrix factorization~\cite{lr3,li2023lossless,wang2024fact}, knowledge distillation~\cite{distw2v,distilhubert,wang2022lighthubert,deepvswide,fithubert,onepass-zq}, and model pruning~\cite{pru3,pru5,gu2024sparsewav,wang2023task,jiang2023accurate,lodagala2023pada,hj,li25v_interspeech,xu25e_interspeech}.

Structured coarse-grained pruning stands out among these as it requires neither special computational operators nor a teacher model.
However, previous studies on model pruning of speech foundation ASR systems face the following limitations: 

\textbf{1) Neglecting the similarity between parameters.} 
Existing importance-based pruning methods~\cite{hj,xu25e_interspeech,peng2023dphubert} evaluate the importance of each component in isolation. Consequently, even when two high-importance weights are functionally redundant, these methods fail to prune either of them.
\textbf{2) Heavy reliance on raw data and fine-tuning.} 
This reliance can evolve into a notable drawback when access to original training data is limited.
Notably, even methods that eliminate the fine-tuning requirement often still necessitate raw data for calibration. This general dependence on data becomes a critical impediment in scenarios where training data are inaccessible.
\textbf{3) 
\tim{Challenges} in hardware-friendly implementation.} 
\tim{Fine-grained unstructured pruning methods ~\cite{ wang2024fact, gu2024sparsewav, xu25e_interspeech, gu2025efficient} typically achieve minimal performance degradation but rely on specialized hardware or libraries for acceleration. This limits their generalizability, particularly for edge devices (e.g., mobile phones), where such hardware or libraries are often unavailable.}

To this end, this paper proposes a novel compression approach for speech foundation models that clusters different structures instead of pruning them. Our method employs a \textbf{structured, coarse-grained compression strategy}, yielding models that are readily deployable and compatible with general-purpose hardware.
To further improve performance, we propose \textit{mixed sparsity}, a strategy where we leverage layer-wise variance to assign different numbers of clusters to each layer for parameter clustering. 
The opposite, applying a single sparsity to all layers, is termed \textit{uniform sparsity}.

All experiments are conducted on HuBERT-\textit{large} and Whisper-\textit{large-v3} using the LibriSpeech~\cite{panayotov2015librispeech} dataset. 
We also fine-tuned HuBERT-\textit{large} to further explore parameter initialization differences between our method and magnitude-based pruning (MP), and extended our mixed sparsity strategy to MP by assigning different numbers of parameter units to retain to each layer based on layer-wise variance.
\textbf{1)} For HuBERT-\textit{large}, our method outperforms MP with uniform sparsity when overall sparsity reaches 30\% or higher. At 50\% uniform sparsity, our data-free method achieves its largest gains over MP, with absolute WER reductions of 27.73\% (34.37\% relative) on test-clean and 18.61\% (21.91\% relative) on test-other. Following fine-tuning, the advantage persists, with our method outperforming MP by 0.19\% (3.36\% relative) and 0.79\% (4.62\% relative) in absolute WER, respectively.
\textbf{2)} The effectiveness of our data-free approach is further demonstrated on Whisper-\textit{large-v3}, where it delivers absolute WER reductions of 2.86\% (test-clean) and 5.02\% (test-other) over MP, all while achieving a 10\% model compression with no significant WER increase across any of the four dev/test subsets relative to the uncompressed baseline.


The main contributions of this paper include:

    \textbf{1)} 
    \zq{Instead of conventional pruning, our approach is based on parameter clustering and fusion. Importance-score-based methods, such as magnitude-based pruning~\cite{han2015learning}, irreversibly discard parameters with low scores, which may be functionally important in concert with others. In contrast, our method identifies and fuses similar components, preserving their collective information to avoid the potential loss of model capability.}
    
    \textbf{2)} 
    \zq{Our approach represents a step toward data-free and training-free compression, which remains largely unexplored in the compression of speech foundation models. While we also present an analysis of fine-tuning HuBERT-\textit{large} to explore its potential benefits, our primary results on Whisper-\textit{large-v3} validate the effectiveness and efficiency of data-free and training-free compression.}

    \textbf{3)} 
    \zq{Our method produces a hardware-friendly, coarse-grained compressed model, which substantially differs from fine-grained, unstructured pruning~\cite{wang2024fact,gu2024sparsewav,xu25e_interspeech,gu2025efficient}. The latter results in irregular sparsity patterns that require specialized software libraries or hardware for efficient inference and often fail to deliver practical speedups.
    By maintaining a structured format, models compressed with our approach can be easily deployed on standard hardware platforms.}
    
\vspace{-3mm}
\section{HuBERT and Whisper Speech Models}
\vspace{-1mm}
Self-supervised learning (SSL) speech models such as HuBERT~\cite{hsu2021HuBERT} and WavLM~\cite{chen2022wavlm}, alongside the weakly-supervised, multi-lingual Whisper~\cite{openai2022whisper}, rely on Transformer backbones that account for the vast majority of their total parameters. Architecturally, HuBERT comprises a CNN feature extractor, a Transformer encoder, a projection layer, and a code embedding layer. Whisper consists of a convolutional input block and a Transformer encoder-decoder. In both architectures, the Transformer blocks comprise a \textbf{M}ulti-\textbf{H}ead \textbf{S}elf-\textbf{A}ttention (MHSA) module and a \textbf{F}eed-\textbf{F}orward \textbf{N}etwork (FFN) module, with Whisper's decoder additionally including a cross-attention module. In this work, we compress the linear layers of the MHSA, cross-attention (${\{\mathbf{W_{q\text{uery}}}, \mathbf{W_{k\text{ey}}}, \mathbf{W_{v\text{alue}}}, \mathbf{W_{out}}\}}$), and FFN (${\{\mathbf{W_{fc1}}}, {\mathbf{W_{fc2}}\}}$) modules within HuBERT-\textit{large}'s encoder and Whisper-\textit{large-v3}'s encoder and decoder blocks.
\vspace{-3mm}
\section{Magnitude-based Pruning}
\vspace{-1mm}
\label{Pre}

Magnitude-based pruning removes parameters based on the principle that those with smaller magnitudes contribute less to the model's performance. 
When applying it to \textbf{structured units} like attention heads or intermediate units, their importance is evaluated by the \textbf{sum of $L_2$-magnitudes} (hereinafter referred to as the \textbf{L2-norm}), where $\lVert{\cdot}\rVert_{2}$ denotes the standard Euclidean norm.
\textbf{1)} For a certain MHSA module as shown in Fig.~\ref{frame}(top), the $m$-th attention head's importance $\mathbf{S}_{\mathrm{MHSA}}$ is calculated as 
{
\setlength{\abovedisplayskip}{3pt}
\setlength{\belowdisplayskip}{3pt}
\begin{equation}
\mathbf{S}_{\mathrm{MHSA}}
 = \sum_{\mathbf{W} \in \{\mathbf{W}_\text{q}, \mathbf{W}_\text{k}, \mathbf{W}_\text{v}, {\mathbf{W}^\top_\text{out}}\}} 
\lVert \mathbf{W}\big|_{m \cdot d_h : (m+1) \cdot d_h,\ :} \rVert_2 ~~,
\label{s1}
\end{equation}
}
where $d_h$ is the head dimension. The cross-attention module follows this identical calculation. \textbf{2)} For a certain FFN module as shown in Fig.~\ref{frame}(bottom left), the $n$-th intermediate unit's importance $\mathbf{S}_{\mathrm{FFN}}$ is calculated as
{
\setlength{\abovedisplayskip}{3pt}
\setlength{\belowdisplayskip}{3pt}
\begin{equation}
\mathbf{S}_{\mathrm{FFN}}
 = \sum_{\mathbf{W} \in \{\mathbf{W}_\text{fc1}, {\mathbf{W}^\top_\text{fc2}}\}} 
\lVert \mathbf{W}\big|_{n,\ :} \rVert_2 ~~.
\label{s2}
\end{equation}
}
Given the global sparsity $sp \in [0, 1)$ representing the percentile of the least important units to be removed, the \textbf{target count} for a specific module is $K = \operatorname{round}(N \times (1-sp))$. Here, $N$ denotes the original number of structured units in that module, yielding a compression ratio of $1/(1-sp)$.



\vspace{-0.3cm}
\section{Parameter Clustering}

\vspace{-1mm}
\subsection{Structured compression using parameter clustering}
\vspace{-2mm}

\label{4.1}
\begin{figure*}[h]
    \centering
    \includegraphics[scale=0.074]{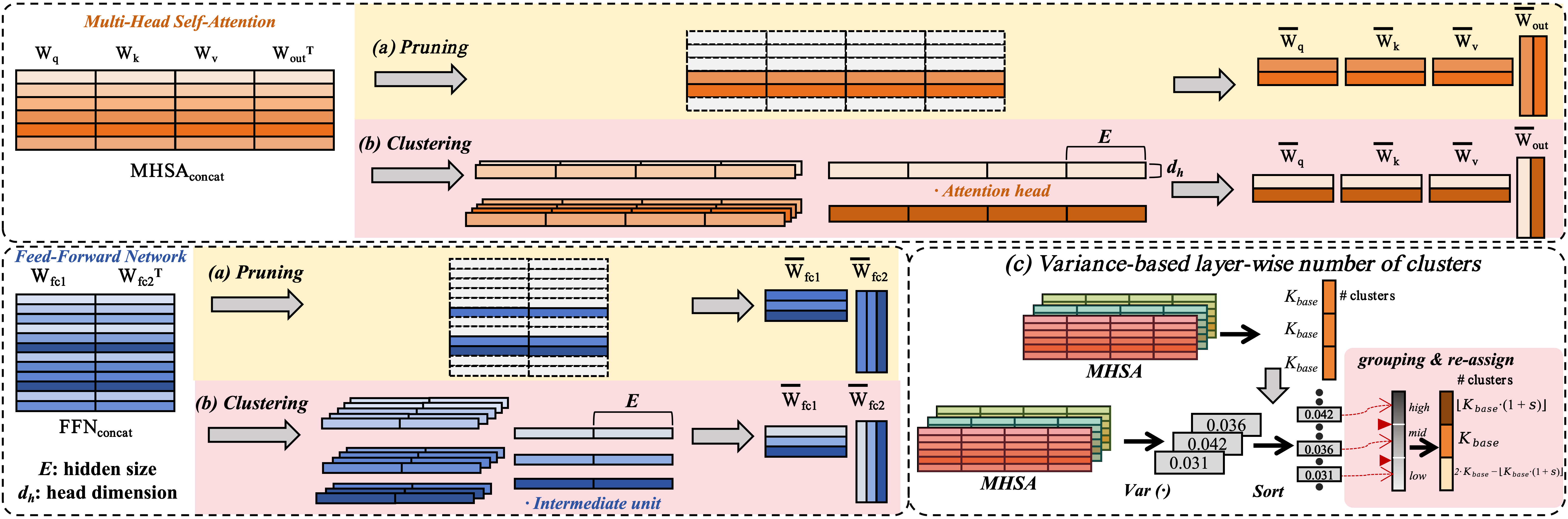}
    \vspace{-3mm}
    \caption{For a certain layer (e.g., target counts $K_l=2$ and $3$ are kept for an Encoder MHSA and FFN module, respectively): \textbf{(a)} Magnitude-based pruning retains top-$K$ structured units via L2-norm; \textbf{(b)} Parameter clustering merges structured units into fewer clusters (e.g., MHSA: 2 units $\to$ one cluster, 4 units $\to$ another cluster); \textbf{(c)} Variance-based mixed sparsity assigns adaptive $K_l$ to modules based on their variance levels (e.g., allocating  $\lfloor 1.2 K_{\text{base}} \rfloor$,                                                                                            $K_{\text{base}}$ and $2\cdot K_{\text{base}}-\lfloor 1.2 K_{\text{base}} \rfloor$ to three distinct MHSA modules).}
    \vspace{-6mm}
    \label{frame}
\end{figure*}

Unlike pruning, which permanently discards parameters, \textbf{parameter clustering} reduces the model size by merging similar structured units within Attention and FFN modules. A key advantage of our approach is its \textbf{data-free and training-free} nature.
For each module, the \textbf{target count} $K = \operatorname{round}(N \times (1-sp))$ defines the number of centroids to be retained, where $sp \in [0, 1)$ is the global sparsity and $N$ is the original number of structured units within that module.

Let $\mathcal{U}_{\text{in}} = \{\mathbf{u}_1, \mathbf{u}_2, \dots, \mathbf{u}_N\}$ denote the set of $N$ candidate structured units. For consistent distance measurement, we apply a flattening operation $\text{vec}(\cdot)$ to convert each unit $\mathbf{u}_i$ into a 1D vector. We use k-means clustering to map $\mathcal{U}_{\text{in}}$ into a compressed set of $K$ centroids $\mathcal{U}_{\text{out}} = \{\mathbf{c}_1, \mathbf{c}_2, \dots, \mathbf{c}_K\}$, obtained via iterative hard assignment and centroid update:
{
\setlength{\abovedisplayskip}{3pt}
\setlength{\belowdisplayskip}{3pt}
\begin{equation}
\mathcal{U}_{\text{out}} = \operatorname{k-means}(\mathcal{U}_{\text{in}}, K).
\end{equation}
}
The optimization objective is to minimize the within-cluster sum of squares (WCSS):
{
\setlength{\abovedisplayskip}{3pt}
\setlength{\belowdisplayskip}{3pt}
\begin{equation}
\sum_{i=1}^{N} \min_{\mathbf{c}_j \in \mathcal{U}_{\text{out}}} \lVert \text{vec}(\mathbf{u}_i) - \mathbf{c}_j \rVert_2^2.
\end{equation}
}
This objective seeks centroids that minimize the total squared error between each unit and its nearest centroid, yielding a locally optimal compression.


As illustrated in Fig.~\ref{frame}, in each MHSA block, the weight matrices $\mathbf{W}_{\text{q}}$, $\mathbf{W}_{\text{k}}$, $\mathbf{W}_{\text{v}}$, and the transpose of $\mathbf{W}_{\text{out}}$, denoted $\mathbf{W}_{\text{out}}^\top$, are concatenated into a matrix $\mathbf{MHSA}_{\text{concat}}$. Here, the set $\mathcal{U}_{\text{in}}$ consists of $N$ attention heads. Since each head is a sub-matrix $\mathbf{u}_i$ with a shape of $d_{h} \times 4E$, where $d_h$ is the head dimension and $E$ is the hidden size, the operation $\text{vec}(\mathbf{u}_i)$ flattens it for distance calculation. Similarly, for the FFN block, $\mathbf{W}_{\text{fc1}}$ and $\mathbf{W}_{\text{fc2}}^\top$ are concatenated into $\mathbf{FFN}_{\text{concat}}$, where $\mathcal{U}_{\text{in}}$ consists of $N$ intermediate units, each with the shape of $1 \times 2E$. Furthermore, the cross-attention module in the decoder follows the exact same concatenation and flattening rules as the MHSA block.
Finally, the compression process is executed by replacing the original units with the optimal centroids $\mathcal{U}_{\text{out}}$. These centroids are stacked to reconstruct the compressed concatenated matrix, which is subsequently split back into its constituent compressed weight matrices (i.e., $\overline{\mathbf{W}}_{\text{q}}$, $\overline{\mathbf{W}}_{\text{k}}$, $\overline{\mathbf{W}}_{\text{v}}$, $\overline{\mathbf{W}}_{\text{out}}$, $\overline{\mathbf{W}}_{\text{fc1}}$ and $\overline{\mathbf{W}}_{\text{fc2}}$).
\vspace{-3mm}
\subsection{Parameter variance-based mixed sparsity allocation}
\label{msp}
\vspace{-2mm}

Intuitively, a high variance of parameters within $\mathbf{MHSA}_{\text{concat}}$ or $\mathbf{FFN}_{\text{concat}}$ suggests that the constituent units vary significantly in information content. We posit that matrices that show higher variance encapsulate more complex information and thus require a larger budget of target units $K$ to preserve performance. Based on this intuition, we propose a \textbf{variance-based mixed sparsity} strategy.

We define the \textbf{module group} as the aggregate of functionally identical modules across all layers, within which sparsity allocation is conducted independently. Specifically: 1) \textbf{HuBERT-\textit{large}:} It forms 2 groups (Encoder MHSA/FFN), each containing 24 modules. 2) \textbf{Whisper-\textit{large-v3}:} It forms 5 groups (Encoder MHSA/FFN, Decoder MHSA/FFN/cross-attention), each containing 32\footnote{For Whisper-\textit{large-v3}, we set $|G_{\text{low}}|=11$, $|G_{\text{mid}}|=10$, and $|G_{\text{high}}|=11$, where $|\cdot|$ denotes the group size.} modules.
Within each module group, the modules are sorted by their parameter variance and then evenly partitioned into three sub-groups:
{
\setlength{\abovedisplayskip}{3pt}
\setlength{\belowdisplayskip}{3pt}
\begin{equation}
    G = \{ G_{\text{low}}, G_{\text{mid}}, G_{\text{high}} \}.
    \label{eq:groups}
\end{equation}
}
For a module in the $l$-th sub-group ($l \in \{\text{low, mid, high}\}$), we assign a specific target count $K_l$ for both pruning and clustering.
Specifically, groups with higher variance are considered more sensitive and are assigned a larger $K_l$ (i.e., subjected to less compression), while groups with lower variance are assigned a smaller $K_l$. This allocation is defined as:
{
\setlength{\abovedisplayskip}{3pt}
\setlength{\belowdisplayskip}{3pt}
\begin{equation}
    K_l = 
    \begin{cases} 
    \lfloor K_{\text{base}} \cdot (1 + s) \rfloor & \text{if } l = \text{high}, \\
    K_{\text{base}} & \text{if } l = \text{mid}, \\
     2 \cdot K_{\text{base}} - \lfloor K_{\text{base}} \cdot (1 + s) \rfloor & \text{if } l = \text{low}.
    \end{cases}
    \label{eq:sparsity_delta}
\end{equation}
}
Here, the hyperparameter $s > 0$ and $K_{\text{base}} = \operatorname{round}(N \times (1-sp))$ represents the uniform target count derived from the global sparsity $sp$. Furthermore, to prevent exceeding the model capacity, if $\lfloor K_{\text{base}} \cdot (1 + s) \rfloor$ surpasses the uncompressed structured unit count $N$, it is capped at $N$. Notably, both mixed and uniform sparsity yield identical overall sparsity, as the former merely reallocates the budget within each module group.


\vspace{-3mm}
\section{Experiments}
\vspace{-2mm}
\subsection{Experimental setup}
\vspace{-0.2cm}
\noindent\textbf{Uncompressed baselines and data.} For HuBERT-\textit{large}, we fine-tuned HuBERT-large-ll60k\footnote{\href{https://huggingface.co/facebook/HuBERT-large-ll60k}{Huggingface: facebook/HuBERT-large-ll60k}} for 20 epochs as our baseline, with other setups consistent with those in \textbf{Post-clustering fine-tuning}.
For Whisper-\textit{large}, we downloaded Whisper-\textit{large-v3}\footnote{\href{https://huggingface.co/openai/whisper-large-v3}{Huggingface: openai/whisper-large-v3}} as our baseline. All systems are evaluated on the LibriSpeech dev and test datasets. 

\noindent\textbf{Post-clustering fine-tuning.}
We also fine-tuned the HuBERT-\textit{large} after data-free clustering on LibriSpeech~\cite{panayotov2015librispeech} 100-hour clean subset for 3 \tim{epochs}.
We utilized the AdamW optimizer with a learning rate of 2e-4 and a batch size of 16. A linear warm-up is implemented for the first 10\% of the training steps, followed by a linear decay to zero. All experiments were conducted on a single NVIDIA A40 (48 GB).

\noindent\textbf{Variance-based mixed sparsity.} 
In all experiments, the hyperparameter $s$ (from Sec.~\ref{msp}) is set to 0.2. We also clarify that "sparsity" in this paper refers to the average sparsity over all the module groups only, excluding the other parts.



\begin{table*}[ht]
\caption{WER ($\downarrow$) Comparison between parameter clustering and magnitude-based pruning (before and after fine-tuning) on HuBERT-\textit{large}. Cells with a blue background indicate that parameter clustering exhibits no statistically significant (MAPSSWE~\cite{gillick1989some}, $\alpha$=0.05) WER increase relative to its magnitude-based pruning counterpart. \textbf{Bold} indicates significant outperformance over the magnitude-based pruning counterpart with its subscript denoting the absolute WER reduction, with red highlighting the maximum reduction within that subset.}
\vspace{-3mm}
    \centering
    \setlength\tabcolsep{7pt}  
    \renewcommand{\arraystretch}{1.0}
    \resizebox{\linewidth}{!}{  
    \begin{tabular}{c|c|c|c|c|c|>{\centering\arraybackslash}p{1.4cm}>
    {\centering\arraybackslash}p{1.4cm}>{\centering\arraybackslash}p{1.4cm}>{\centering\arraybackslash}p{1.4cm}|>{\centering\arraybackslash}p{1.4cm}>
{\centering\arraybackslash}p{1.4cm}>{\centering\arraybackslash}p{1.4cm}>{\centering\arraybackslash}p{1.4cm}}
    \hline
    \hline
    \multirow{2}{*}{ID} & \multirow{2}{*}{Sparsity} & \#Params  & \multirow{2}{*}{\shortstack{Compression\\Method}} & \multirow{2}{*}{\shortstack{Mixed\\sparsity}} & \multirow{2}{*}{\shortstack{GFLOPs\\Total / Trans.}}
    & \multicolumn{4}{c|}{\shortstack{Before Fine-tuning}}  
    & \multicolumn{4}{c}{\shortstack{After Fine-tuning}} \\  
    &  & (Millions) & & & & dev-clean & dev-other & test-clean & test-other  
 & dev-clean & dev-other & test-clean & test-other \\  
    \hline
    0 & \multicolumn{4}{c|}{Uncompressed HuBERT Large (316.6M)}  
    & 36.25 / 30.44 & 3.35 & 8.13 & 3.44 & 8.34  
     & 3.35 & 8.13 & 3.44 & 8.34 \\  
    \hline
    \hline
    1 & \multirow{4}{*}{10\%}  & \multirow{4}{*}{282.7M} & \multirow{2}{*}{Magnitude} & \xmark 
    & \multirow{4}{*}{32.94 / 27.14} & 4.24 & 11.16 & 4.40 & 11.56  
    & 3.78 & 10.29 & 3.83 & 10.39 \\  
    2 &  & &  & \cmark 
    & & 3.63 & 10.46 & 3.77 & 10.64  
    & 3.62 & 10.18 & 3.75 & 10.04 \\  
    \cline{4-5}  
    3 &  & & \multirow{2}{*}{clustering} & \xmark 
    & & \textbf{4.13}\abswer{ 0.11} & 11.42 & {\cellcolor{cyan!15}4.29} & {\cellcolor{cyan!15}11.60}  
    & {\cellcolor{cyan!15}3.71} & {\cellcolor{cyan!15}10.21} & {\cellcolor{cyan!15}3.92} & {\cellcolor{cyan!15}10.19} \\  
    4 &  & &  & \cmark 
    & & {\cellcolor{cyan!15}3.63} & 11.00 & {\cellcolor{cyan!15}3.80} & 11.01  
    &  {\cellcolor{cyan!15}3.58} &  {\cellcolor{cyan!15}10.42} &  {\cellcolor{cyan!15}3.77} &  {\cellcolor{cyan!15}10.28} \\  
    \hline
    5 & \multirow{4}{*}{20\%}  & \multirow{4}{*}{256.3M} & \multirow{2}{*}{Magnitude} & \xmark 
    &\multirow{4}{*}{30.29 / 24.49} & 6.36 & 15.67 & 6.48 & 15.96  
    & 3.86 & 10.83 & 4.10 & 11.05 \\  
    6 &  & &  & \cmark 
    & & 4.08 & 13.09 & 4.39 & 13.56  
    & 3.69 & 10.40 & 3.87 & 10.54 \\  
    \cline{4-5}
    7 &  & & \multirow{2}{*}{clustering} & \xmark 
    & & {\cellcolor{cyan!15}6.28} & 16.16 & {\cellcolor{cyan!15}6.42} & 16.58  
    & {\cellcolor{cyan!15}3.94} & 11.29 & {\cellcolor{cyan!15}4.04} & {\cellcolor{cyan!15}11.20} \\  
    8 &  & &  & \cmark &
    & {\cellcolor{cyan!15}4.10} & 14.34 & {\cellcolor{cyan!15}4.31} & 14.92  
    & {\cellcolor{cyan!15}3.80} & 10.74 & {\cellcolor{cyan!15}3.88} & 11.07 \\  
    \hline
    9 & \multirow{4}{*}{30\%}  & \multirow{4}{*}{223.6M} & \multirow{2}{*}{Magnitude} & \xmark 
    &\multirow{4}{*}{26.98 / 21.18} & 16.07 & 29.00 & 16.40 & 29.37  
    & 4.15 & 12.55 & 4.44 & 12.99 \\  
    10 &  & &  & \cmark 
    & & 7.16 & 21.37 & 7.16 & 22.52  
    & 3.96 & 11.86 & 4.12 & 12.06 \\  
    \cline{4-5}
    11 &  & & \multirow{2}{*}{clustering} & \xmark 
    & & \textbf{13.92}\abswer{ 2.15} & \textbf{26.66}\abswer{ 2.34} & \textbf{13.99}\abswer{ 2.41} & \textbf{27.13}\abswer{ 2.24}  
    & {\cellcolor{cyan!15}4.28} & \textbf{12.02}\abswer{ 0.53} & {\cellcolor{cyan!15}4.44} & \textbf{12.33}\abswer{ 0.66} \\  
    12 &  & &  & \cmark &
    & \textbf{6.89}\abswer{ 0.27} & \textbf{20.14}\abswer{ 1.23} & \textbf{6.90}\abswer{ 0.26} & \textbf{21.00}\abswer{ 1.52}  
    & 4.09 & {\cellcolor{cyan!15}11.98} & {\cellcolor{cyan!15}4.22} & {\cellcolor{cyan!15}11.97} \\  
    \hline
    13 & \multirow{4}{*}{40\%}  & \multirow{4}{*}{197.1M} & \multirow{2}{*}{Magnitude} & \xmark 
    &\multirow{4}{*}{24.33 / 18.53} & 37.26 & 48.48 & 37.58 & 48.37  
    & 4.60 & 13.26 & 4.72 & 13.53 \\  
    14 &  & &  & \cmark 
    & & 18.02 & 36.02 & 18.26 & 37.12  
    & 4.30 & 12.79 & 4.42 & 13.17 \\  
    \cline{4-5}
    15 &  & & \multirow{2}{*}{clustering} & \xmark &
    &\textbf{32.25}\abswer{ 5.01} &	\textbf{46.90}\abswer{ 1.58} &	\textbf{32.54}\abswer{ 5.04} &	\textbf{47.12}\abswer{ 1.25}
    &{\cellcolor{cyan!15}4.55} 	&13.88 &	{\cellcolor{cyan!15}4.78}& 	14.23  \\  
    16 &  & &  & \cmark & 
    &\textbf{15.95}\abswer{ 2.07} &	\textbf{33.33}\abswer{ 2.69} &	\textbf{15.87}\abswer{ 2.39} 	&\textbf{34.55}\abswer{ 2.57}   
    &{\cellcolor{cyan!15}4.38} &	13.69 &	{\cellcolor{cyan!15}4.48} &	13.62 \\  

    \hline
    17 & \multirow{4}{*}{50\%}  & \multirow{4}{*}{164.4M} & \multirow{2}{*}{Magnitude} & \xmark 
    & \multirow{4}{*}{21.03 / 15.22} & 81.23 & 85.40 & 80.68 & 84.92  
    & 5.28 & 16.86 & 5.65 & 17.09 \\  
    18 &  & &  & \cmark 
    &  & 51.33 & 67.00 & 50.62 & 67.24  
    & 4.85 & 16.38 & 5.10 & 16.44 \\  
    \cline{4-5}
    19 &  & & \multirow{2}{*}{clustering} & \xmark 
    & & \textbf{52.92}\abswerblue{ 28.31} & \textbf{66.07}\abswerblue{ 19.33} & \textbf{52.95}\abswerblue{ 27.73} & \textbf{66.31}\abswerblue{ 18.61}  
    & {\cellcolor{cyan!15}5.41} & \textbf{16.09}\abswer{ 0.77} & \textbf{5.46}\abswerblue{ 0.19} & \textbf{16.30}\abswer{ 0.79} \\  
    20 &  & &  & \cmark 
    & & \textbf{41.94}\abswer{ 9.39} & \textbf{60.66}\abswer{ 6.34} & \textbf{41.10}\abswer{ 9.52} & \textbf{61.48}\abswer{ 5.76}  
    & {\cellcolor{cyan!15}5.00} & \textbf{15.54}\abswerblue{ 0.84} & {\cellcolor{cyan!15}5.12} & \textbf{15.68}\abswer{ 0.76} \\  
    \hline
    21 & \multirow{4}{*}{60\%}  & \multirow{4}{*}{131.6M} & \multirow{2}{*}{Magnitude} & \xmark &\multirow{4}{*}{17.72 / 11.92} & 87.28 & 91.49 & 86.43 & 91.27  
    & 6.81 & 20.37 & 6.99 & 20.98 \\  
    22 &  & &  & \cmark 
    & & 88.99 & 93.21 & 88.10 & 92.96  
    & 6.40 & 19.74 & 6.59 & 20.17 \\  
    \cline{4-5}
    23 &  & & \multirow{2}{*}{clustering} & \xmark 
    & & \textbf{61.15}\abswer{ 26.13} & \textbf{75.78}\abswer{ 15.71} & \textbf{60.69}\abswer{ 25.74} & \textbf{76.12}\abswer{ 15.15}  
    & {\cellcolor{cyan!15}6.63} & \textbf{19.60}\abswer{ 0.77} & {\cellcolor{cyan!15}6.83} & \textbf{19.76}\abswerblue{ 1.22} \\  
    24 &  & &  & \cmark 
    & & \textbf{76.42}\abswer{ 12.57} & \textbf{87.63}\abswer{ 5.58} & \textbf{76.28}\abswer{ 11.82} & \textbf{87.35}\abswer{ 5.61}  
    & \textbf{6.21}\abswerblue{0.19} & \textbf{19.17}\abswer{ 0.57} & {\cellcolor{cyan!15}6.43} & \textbf{19.52}\abswer{ 0.65} \\  
    \hline
    \hline
    \end{tabular} }
    \vspace{-6mm}
    \label{tab1}
\end{table*}

\begin{table}[h]
\caption{WER ($\downarrow$) Comparison between parameter clustering and magnitude-based pruning on Whisper-\textit{large-v3}. ${\dag}$ indicates no statistically significant (MAPSSWE~\cite{gillick1989some}, $\alpha$=0.05) WER increase relative to the uncompressed baseline. The marks have the same meaning as those in Table~\ref{tab1}.}
\vspace{-3mm}
    \centering
    \setlength\tabcolsep{1pt}
    \resizebox{\linewidth}{!}{
    \begin{tabular}{c|c|c|c|c|c|cccc}
    \hline
    \hline
    \multirow{2}{*}{ID} & \multirow{2}{*}{Sparsity} & \#Params  & \multirow{2}{*}{\shortstack{Compression\\Method}} & \multirow{2}{*}{\shortstack{Mixed\\sparsity}} & \multirow{2}{*}{\shortstack{GFLOPs\\Total / Trans.}} & \multirow{2}{*}{dev-clean} &\multirow{2}{*}{dev-other} &\multirow{2}{*}{test-clean} &\multirow{2}{*}{test-other}\\
        &  &Millions & & & & & &\\
        \hline
        \multirow{2}{*}{0} & 
        \multicolumn{4}{c|}{\multirow{2}{*}{Uncompressed Whisper-\textit{large-v3} (1550M)}} &
        \multirow{2}{*}{\shortstack{2773.53 \\/ 2743.66}} 
         &
        \multirow{2}{*}{ 2.06} &
        \multirow{2}{*}{ 4.25} &
        \multirow{2}{*}{ 2.03} &
        \multirow{2}{*}{ 3.99} \\
        & \multicolumn{4}{c|}{} & & & & & \\
        \hline
        \hline
        1 & \multirow{4}{*}{10\%}  & \multirow{4}{*}{1397M} & \multirow{2}{*}{Magnitude} &\xmark &\multirow{4}{*}{\shortstack{2499.16\\ / 2469.30}} &66.29 &	72.30& 	69.79 &	75.68   \\
        2 &  &  &  &\cmark  & & 5.00 &	7.15 	&4.83 &	9.08   \\
        \cline{4-5}
        3 &  &  & \multirow{2}{*}{clustering} &\xmark  & &\textbf{2.89}\abswer{63.40} &	\textbf{5.89}\abswer{66.41} &	\textbf{2.34}\abswer{67.45} &	\textbf{7.10}\abswer{68.58}  \\
        4 &  &  &  &\cmark &  &\textbf{1.89}$^{\dag}$\abswer{3.11} &\textbf{4.41}$^{\dag}$\abswer{2.74} 	&\textbf{1.97}$^{\dag}$\abswer{2.86} &	\textbf{4.06}$^{\dag}$\abswer{5.02}   \\
        \hline
        5 & \multirow{4}{*}{20\%}   & \multirow{4}{*}{1250M} & \multirow{2}{*}{Magnitude} &\xmark &\multirow{4}{*}{\shortstack{2224.80\\ / 2194.93}} &100.00 &	100.00 &	100.00 &	100.00   \\
        6 &  &  &  &\cmark & &124.40 &	121.86 &	116.57 &	126.49   \\
        \cline{4-5}
        7 &  &  & \multirow{2}{*}{clustering} &\xmark  & &\textbf{13.59}\abswer{86.41} &	\textbf{20.85}\abswer{79.15}& 	\textbf{15.31}\abswer{84.69} &	\textbf{24.94}\abswer{75.06}  \\
        8 &  &  &  &\cmark & &\textbf{3.74}\abswerblue{120.66} &	\textbf{7.77}\abswerblue{114.09} 	&\textbf{3.65}\abswerblue{112.92} &	\textbf{6.65}\abswerblue{119.84} \\
        \hline
        \hline
    \end{tabular} }
    \label{tab2}
    \vspace{-8mm}
\end{table}

\vspace{-0.3cm}
\subsection{Main results}
\vspace{-0.1cm}
\subsubsection{Data-free clustering before fine-tuning}
\vspace{-0.1cm}

\textbf{1) \tim{Comparison} with Magnitude-based Pruning (MP):} As shown in Tab.~\ref{tab1}, for HuBERT-\textit{large} at uniform sparsity of 30\% or higher, our method outperforms MP on all subsets (e.g., ID~11 vs. ID~9). An average absolute reduction in WER on all subsets of 23.50\% is observed against MP at 50\% sparsity (ID~19 vs. ID~17).

For Whisper-\textit{large-v3} shown in Tab.~\ref{tab2}, our method significantly reduces the WER on test-clean and test-other compared to MP at 10\% sparsity, achieving absolute reductions of 67.45\% and 68.58\% (ID~3 vs. ID~1), respectively. With mixed sparsity, our method still delivers substantial absolute WER reductions of 2.86\% and 5.02\% over MP (ID~4 vs. ID~2). In particular, our method compresses Whisper-\textit{large-v3} with no statistically significant WER increase on any subset against the uncompressed baseline (ID~4 vs. ID~0). At a higher mixed sparsity of 20\%, our method's WER increases by only 1.62\% and 2.66\% on the test-clean and test-other subsets over the uncompressed baseline (ID~8 vs. ID~0), respectively, whereas MP suffers catastrophic degradation across all subsets (ID~5, ID~6 vs. ID~0).

\textbf{2) Comparison between uniform and mixed sparsity:}
Furthermore, for HuBERT-\textit{large} shown in Tab.~\ref{tab1}, the mixed sparsity strategy improves the performance of the compressed model across the sparsity range from 10\% to 50\% (e.g., ID ~10 vs. ID~9; ID~12 vs. ID~11). However, a performance degradation is observed at 60\% sparsity for both MP and our method. We hypothesize that variance-based mixed sparsity strategy is no longer sufficient to help the model retain critical parameters at extreme sparsity.

For Whisper-\textit{large-v3} in Tab.~\ref{tab2}, compared to uniform sparsity, the mixed sparsity approach at 20\% sparsity yields WER reductions of 11.66\% absolute (76.16\% relative) on test-clean and 18.29\% absolute (73.34\% relative) on test-other (ID~8 vs. ID~7). However, it should be noted that once the sparsity exceeds 20\% (such as 30\%), our method also suffers catastrophic degradation, regardless of whether mixed sparsity is used.

\vspace{-3mm}
\subsubsection{Post-clustering fine-tuning}
\vspace{-0.1cm}
\textbf{1) \tim{Comparison} with magnitude-based pruning (MP):} As shown in Tab.~\ref{tab1}, at sparsity of 50\% or higher, fine-tuned HuBERT-\textit{large} with our method significantly outperforms MP on the two \textit{other} subsets, while performing on par with or better than MP on the two \textit{clean} subsets (e.g., ID~23 vs. ID~21; ID~24 vs. ID~22). 
Our method achieves absolute WER reductions of up to 0.19\% (3.36\% relative) and 1.22\% (5.82\% relative) on the test-clean (ID~19 vs. ID~17) and test-other (ID~23 vs. ID~21) subsets, respectively. This shows that our method provides a superior starting point for fine-tuning at extreme sparsity.

\textbf{2) Comparison between uniform and mixed sparsity:}
For HuBERT-\textit{large} in Tab.~\ref{tab1}, at sparsity of 20\% or higher, the models with mixed sparsity consistently outperform their uniform sparsity counterparts at all sparsity \tim{levels} after fine-tuning, regardless of whether our method or MP is used (e.g., ID~22 vs. ID~21; ID~24 vs. ID~23). This further validates the feasibility of our variance-based mixed sparsity pruning strategy.
\vspace{-3mm}
\subsubsection{Hardware Acceleration Analysis}
\vspace{-1mm}
The GFLOPs (giga floating-point operations) of HuBERT's encoder-only design depend solely on the input audio length, set to 1s in our evaluation. Conversely, Whisper's autoregressive Seq2Seq structure requires a predefined text generation length, set to $100$ in our experiments, as the decoder's cross-attention accumulates computational overhead per generated token. As shown in Tab.~\ref{tab1} for HuBERT-\textit{large} and Tab.~\ref{tab2} for Whisper-\textit{large-v3}, both models exhibit a shared trend under structured compression. Reducing the number of attention heads and FFN dimensions leads to a proportional decrease in ``Transformer-only GFLOPs''. However, the ``Total System GFLOPs'' decline at a much slower rate because static uncompressed components, such as the CNN feature extractor, remain unchanged.


\vspace{-3mm}
\subsubsection{Variance Perspective on Compression Sensitivity}
\vspace{-1mm}

The previous experimental results indicate that magnitude-based pruning performs far better on HuBERT-\textit{large} than on Whisper-\textit{large-v3}. 
As detailed in Tab.~\ref{tab:variance_comp}, the overall mean variance and the variance range are derived from the layer-wise variances. HuBERT-\textit{large} exhibits significantly higher overall mean variance compared to substantially smaller variances observed in the five module groups of Whisper-\textit{large-v3}. This discrepancy explains the performance gap. For a model with smaller layer-wise variances, the importance of individual parameters cannot be easily separated based solely on their magnitude. Thus, directly pruning weights with smaller magnitudes is much more likely to cause severe performance degradation.
\begin{table}[h]
    \vspace{-3mm}
    \caption{Layer-wise variance comparison between HuBERT-\textit{large} and Whisper-\textit{large-v3}.}
    \vspace{-3mm}
    \centering
    \setlength\tabcolsep{7pt}
    \resizebox{\linewidth}{!}{
    \begin{tabular}{c|c|c|c}
    \hline
    \hline
    \multirow{2}{*}{Model} & \multirow{2}{*}{\shortstack{Module Group}} & \multirow{2}{*}{\shortstack{Overall mean}} & \multirow{2}{*}{Range (min/max)} \\
    & & & \\
    \hline
    \multirow{2}{*}{HuBERT-\textit{large}} & Encoder MHSA & $1.5\times10^{-2}$ & $[0.7, 2.4]\times10^{-2}$\\
    & Encoder FFN & $1.8\times10^{-2}$ & $[1.3, 2.2]\times10^{-2}$\\
    \hline
    \multirow{5}{*}{Whisper-\textit{large-v3}} & Encoder MHSA & $4.3\times10^{-4}$ & $[3.6, 5.1]\times10^{-4}$\\
    & Encoder FFN & $2.9\times10^{-4}$ & $[2.3, 3.8]\times10^{-4}$\\
    & Decoder MHSA & $4.8\times10^{-4}$ & $[2.3, 9.7]\times10^{-4}$\\
    & Decoder FFN & $3.0\times10^{-4}$ & $[2.6, 3.8]\times10^{-4}$\\
    & Decoder cross-attention & $3.4\times10^{-4}$ & $[0.9, 5.8]\times10^{-4}$\\
    \hline
    \hline
    \end{tabular} }
    \label{tab:variance_comp}
    \vspace{-6mm}
\end{table}
\vspace{-1mm}
\subsubsection{Comparative Analysis of different methods via L2-Norm}
\vspace{-1mm}
For HuBERT-\textit{large}, we first compute the layer-wise averaged L2-norms across all attention heads or FFN intermediate units to obtain exactly 24 layer-wise values. The overall mean and range are then derived directly from these 24 averages. As shown in Tab.~\ref{tab:l2_norm}, at uniform sparsity of 30\%, both magnitude-based pruning and clustering increase the overall L2-norm compared to the uncompressed baseline. However, this increase is notably smaller for our better-performing clustering.

\begin{table}[h]
\vspace{-3mm}
    \caption{Layer-wise averaged L2-norm comparison for HuBERT-Large at uniform sparsity of 30\% before fine-tuning.}
    \vspace{-3mm}
    \centering
    \setlength\tabcolsep{7pt}
    \resizebox{\linewidth}{!}{
    \begin{tabular}{c|c|c|c}
    \hline
    \hline
    \multirow{2}{*}{\shortstack{Module\\Group}} & \multirow{2}{*}{Method} & \multirow{2}{*}{Overall mean} & \multirow{2}{*}{Range (min/max)} \\
    & & & \\
    \hline
    \multirow{3}{*}{Encoder MHSA} & Uncompressed & 3967.94 & [1707.80, 6139.81] \\
    & Pruning & 4606.92 & [2159.27, 7396.73] \\
    & Clustering (Ours) & 4305.46 & [2011.86, 6681.39] \\
    \hline
    \multirow{3}{*}{Encoder FFN} & Uncompressed & 35.27 & [25.19, 42.61] \\
    & Pruning & 46.45 & [33.18, 57.03] \\
    & Clustering (Ours) & 39.69 & [27.96, 50.25] \\
    \hline
    \hline
    \end{tabular} }
    \label{tab:l2_norm}
    \vspace{-3mm}
\end{table}

    
\vspace{-3mm}
\section{Conclusion}
\vspace{-1mm}
We introduce a novel compression method for speech foundation models that utilizes parameter clustering as a data-free and training-free alternative to pruning. A variance-based strategy to re-assign layer-wise sparsity is also explored. Experimental results demonstrate that our method outperforms magnitude-based pruning and achieves results comparable to the baseline.

\section{Generative AI Use Disclosure}
During the preparation of this manuscript, the authors used generative AI tools solely to edit the language and polish the manuscript for better readability. These tools were not used to generate core scientific ideas, experimental data, or technical contributions. All authors have thoroughly reviewed and approved the final manuscript and assume full responsibility for the integrity of its entire content.

\section{Acknowledgements}
This research is supported by Hong Kong RGC GRF grant No. 14200021 and 14200324.

\bibliographystyle{IEEEtran}
\bibliography{refs}

\end{document}